\documentclass[floatfix,pra,twocolumn,showpacs,amsmath,amssymb,letterpaper,groupaddresses,superscriptaddress]{revtex4}

\usepackage{graphicx}
\usepackage{hyperref}
\usepackage{xcolor}
\usepackage{mathtools}
\usepackage{times}
\usepackage{latexsym}
\usepackage{graphicx}
\usepackage{amsmath,amssymb}
\usepackage{verbatim,times,bbm}
\usepackage{placeins}
\usepackage{color}
\usepackage[english]{babel}
\usepackage[T1]{fontenc}
\usepackage{amsmath, amsthm, amssymb, mathrsfs}
\usepackage{epstopdf}
\usepackage{subfigure}
\usepackage{appendix}
\usepackage{pdfpages}

\usepackage{placeins}
\usepackage{float}

\usepackage{times}
\usepackage{latexsym}
\usepackage{graphicx}
\usepackage{verbatim,times,bbm}
\usepackage{color}
\usepackage{appendix}

\newcommand{\be}{\begin{equation}}
\newcommand{\ee}{\end{equation}}

\newcommand{\bea}{\begin{eqnarray}}
\newcommand{\eea}{\end{eqnarray}}

\newcommand{\ba}{\begin{array}}
\newcommand{\ea}{\end{array}}

\newcommand{\bn}{\begin{enumerate}}
\newcommand{\en}{\end{enumerate}}

\usepackage{color}

\begin{document}

\title{Homological analysis of multi-qubit entanglement}

\author{Alessandra Di Pierro\footnote{email: alessandra.dipierro@univr.it}}
\affiliation{Department of Informatics, University fo Verona, Strada Le Grazie 15, 37134 Verona, Italy}

\author{Stefano Mancini\footnote{email: stefano.mancini@unicam.it}}
\affiliation{School of Science and Technology, University of Camerino, I-62032 Camerino, Italy}
\affiliation{INFN-Sezione di Perugia, I-06123 Perugia, Italy}

\author{Laleh Memarzadeh\footnote{email: memarzadeh@sharif.edu}}
\affiliation{Department of Physics, Sharif University of Technology, Teheran, Iran}

\author{Riccardo Mengoni\footnote{email: riccardo.mengoni@univr.it}}
\affiliation{Department of Informatics, University fo Verona, Strada Le Grazie 15, 37134 Verona, Italy}

\begin{abstract}
We propose the usage of persistent homologies to characterize multipartite entanglement. 
On a multi-qubit data set we introduce metric-like measures defined  in terms of bipartite entanglement and then we derive barcodes. { We show that, depending on the distance,  they are able to produce different classifications. In one case, it is possible to obtain the standard separability classes. In the other case, a new  classification of entangled states of three and four qubits  is provided.}
\end{abstract}


\maketitle

\section{Introduction}

In the last few decades the interest on quantum entanglement has been turned from purely foundational/philosophical aspects to more practical/applicative ones, thanks to quantum information processing.
In this setting, the characterization of entanglement is of uppermost importance, although it results a daunting task in multi-partite systems. Many approaches have been developed, based on combinatorics, group theory, geometry, etc. (see e.g. \cite{Horo} for a review).
{However, as far as we know, topological approaches have not been taken into consideration up to now}. 
 
Persistent homology is nowadays widely used to analyse classical data sets represented in the form of \textit{point cloud} \cite{Carl}.
It is a particular sampling based technique from algebraic topology, originally
introduced in \cite{Edel}, aiming at extracting 
topological information from high dimensional data sets. 
 
A quantum approach to computing persistent homology has been devised in \cite{Seth} in order to achieve more efficient algorithms for classical data analysis, but vice versa the usage of persistent homologies in {the study of} quantum data  has not been investigated up to now. Here we propose persistent homologies for characterizing multipartite entanglement. 
On a multi-qubit data set { we introduce semi-metrics defined in terms of bipartite entanglement. We then construct on   the quantum data set  a family of simplicial complexes indexed by a proximity parameter and derive a characterisation of the associated persistent homology by means of barcodes (i.e. a  parameterised version of  Betti numbers). We consider two different notions of distance and show that  they are able to produce different classifications. In one case, it is possible to obtain the known  grouping of separability classes. In the other case, a new  classification of entangled states of three and four qubits  is provided.}

\section{Persistent homology in a nutshell}
Information about topological properties of a topological space $X$, such as connected components, holes, voids etc., are encoded in the homology groups $ \left\lbrace H_{0}(X),H_{1}(X),H_{2}(X),\ldots \right\rbrace $ of the space $X$, where the $ k^{\rm {th}} $ homology group  $ H_{k}(X) $ describes the k-dimensional holes in $ X $ \cite{Hatcher}. 

In order to capture the global topological features in a data set, the corresponding space must first be represented as a simplicial complex, { i.e. as a collection of simple polytopes called simplices \cite{Edel}.}
There are different ways for assigning such simplicial complexes to the data points but all methods are based on distances $\epsilon$ between pairs of points. { For a fixed}  method, by changing $\epsilon$, different topological features (such as connected components, tunnels, voids etc.) { can be}  observed in the global complex. Hence varying $\epsilon$ from small values to sufficiently large ones enables us to find which topological features persist and hence { constitute important properties for the given data set}.  This method of computing multi-scale homological features of the data points is called \textit{persistent homology}. Those features which { at some point} vanish by changing the parameter $\epsilon$ are considered as noise with no particular significance. 

Here we focus on the Rips complex\footnote{Most commonly known as Vietoris-Rips complex, here we  will  call it Rips complex for the sake of brevity.}
to construct simplicial complexes from data points \cite{Carl2}. 
In the Rips complex, $k$-simplices  correspond to $(k + 1)$ points  which are 
pairwise within distance $ \epsilon $. { By} computing topological features for { $\epsilon \in \left( 0,\infty\right) $,} we { can} produce a barcode, i.e.  a collection of horizontal lines in a plane where
the horizontal axis represents $ \epsilon$, while on the vertical axis the  homology generators 
$H_k$ are placed in  arbitrary order.
 Hereafter, a black line in the barcode will indicate a connected component (homology group $H_0 $), a red line will correspond to a hole (homology group $ H_1 $) and a blue line will represent a void (homology group $ H_2 $).

\section{Semi-metrics on quantum data set}

 Consider a quantum dataset $\cal Q$  representing a quantum state $\left|\Psi_n \right\rangle$ over $n$ qubit as a point cloud. This dataset is such that to each point in $\cal Q$ is associated a  single qubit. Let $E(i,j)$ be an entanglement monotone \footnote{{Different  monotones would produce barcodes with the same homology groups generators, up to  sliding or stretching/contraction of the bars.} } between qubit $i$ and $j$. Our aim is to define a { distance} on $\cal Q$ in such a way that the more entangled two qubit are, the closer they are with respect to that { distance}. This naturally leads to define such a distance as:
\begin{equation}
\label{D}
D(i,j):=\left[E(i,j)\right]^{-1},
\end{equation}
with $D(i,j)=0$ iff $i=j$. 
Note that this does not define  a proper metric but a semi-metric because the triangle inequality does not hold. 

 Since $D(i,j)$ only takes into account bipartite entanglement between pairs of qubits and no other form of  entanglement, we introduce a variant of $D(i,j)$  that includes  bipartite entanglement between any possible pair of subsets   containing qubit $i$ and $j$ as: {
\begin{equation}\label{Dtilde}
\widetilde{D}(i,j):= \left[ E(i,j)+\prod_{S\in \mathcal{P}}E(S\cup\{i\},\bar{S}\cup\{j\})\right] ^{-1},
\end{equation} 
where $ S $ and its complement  $ \bar{S} $ belong to the power set $\mathcal{P}$ of the set of all qubit indices except $ i $ and $j$, i.e. the power set of $\{1,2,...,n\}\setminus\{ i,j\}$.

It is clear that also $\widetilde{D}(i,j)$ is a semi-metric because, likewise $D(i,j)$, it does not satisfy the triangle inequality.}

\section{Classification of three qubit states using $\widetilde{D}(i,j)$}

For three qubits, Equation~\eqref{Dtilde} becomes
\begin{equation}
\label{Dtilde3}
\widetilde{D}(i,j)=\Big[E(i,j)+ E(i,\{j,k\}) E(j,\{i,k\}) \Big]^{-1}
\end{equation}
with $i,\neq j\neq k$. {By using this distance, we are able to generate barcodes representing  generators of homology groups $H_0$. This allows us  to distinguish   the following classes of states, reproducing the standard separability classes.}\\

\begin{figure}[htbp]
	\centerline{\includegraphics[scale=0.42]{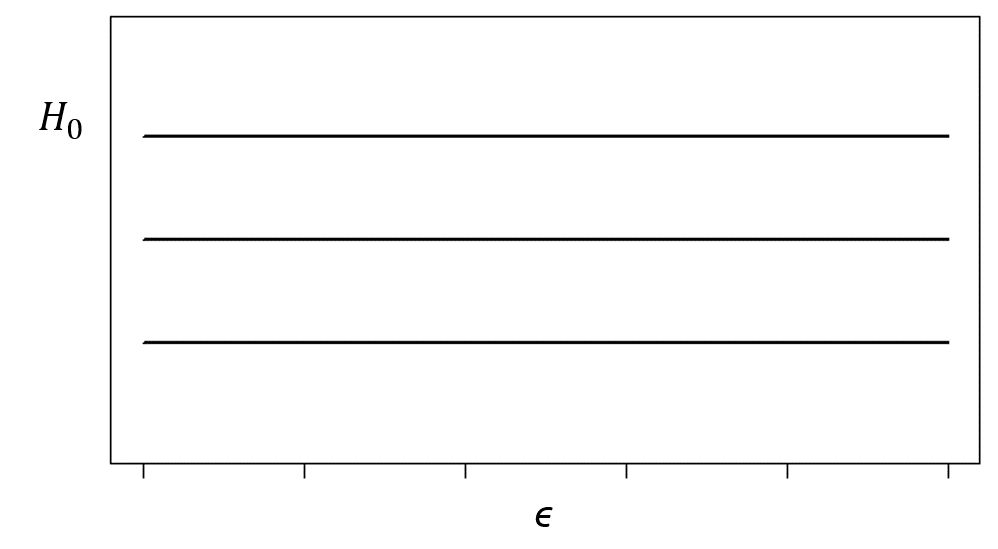}} \caption{Barcode for three qubit { product states}.  }
	\label{figure:1}
\end{figure}

\noindent\emph{{ $\triangleright  $ Product states }}\\
Pure states of this kind are not entangled  and can be written as $|\psi_1\rangle|\psi_2\rangle|\psi_3\rangle$ where $|\psi_i\rangle$ is a generic state for qubit $i$.
In this case, whatever labelling $i,j,k$ of the three qubits, we get $ E(i,j)=0$   $ \forall i,j $ and  
$ E(i,\{j,k\}) = E(j,\{i,k\})=0$. This means that $\widetilde{D}(i,j) =\infty$ for every pair of qubits.
The point cloud associated with this state is  made up of three points placed at infinite distance  from each other, hence the corresponding barcode has the form in Figure~\ref{figure:1}.
\FloatBarrier

\begin{figure}[htbp]
\centerline{\includegraphics[scale=0.42]{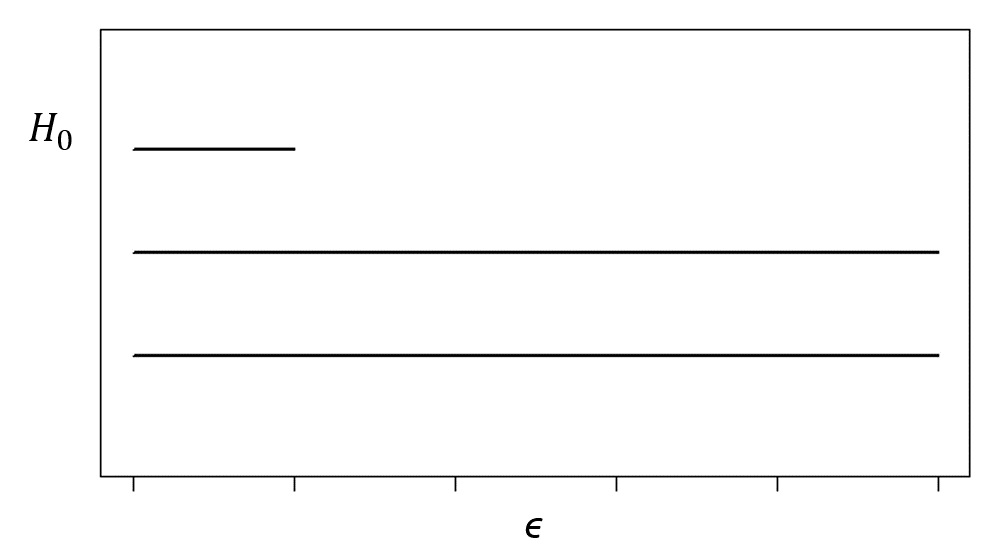}} \caption{Barcode for three qubit bi-separable states. }
\label{figure:2}
\end{figure}

\noindent\emph{{ $\triangleright  $ Bi-separable  states }}\\
Pure states of this kind can be written as  $|\psi_i\rangle|\psi_{jk}\rangle$ where   $|\psi_{jk}\rangle$ is an entangled states between  qubit $ j $ and $ k $.
In this case we get $E(i,j)=E(i,k)=0$ and 
$E(j,k)\neq 0$, while $E(i,\{j,k\})=0$ but $E(j,\{i,k\})\neq 0$ as well as $E(k,\{i,j\})\neq 0$.
This means that $\widetilde{D}(i,j)=\widetilde{D}(i,k)=\infty$ but $\widetilde{D}(j,k) <\infty$. 
Hence the resulting barcode is as in Figure~\ref{figure:2}.
\FloatBarrier
\noindent\emph{{ $\triangleright  $ Fully inseparable states}}\\
Pure states of this kind can only be written as $|\psi_{123}\rangle$ i.e. as an entangled three qubit state.
{ States belonging to this class might have  $ E(i,j)=0$ or  $ E(i,j)\neq 0$  for some or all   pairs of qubits $ i $ and $j $. However  the bipartite entanglement  $E(i,\{j,k\})$  and $E(j,\{i,k\})$ always differ from zero $ \forall i,j,k $.
This means that $ 0<\widetilde{D}(i,j)<\infty $, $ \forall i,j$. The three points in the point cloud are hence  at finite distance  from each other and will connect to form a 2-simplex for sufficiently large $ \epsilon $.}
\begin{figure}[htbp]
	\centerline{\includegraphics[scale=0.42]{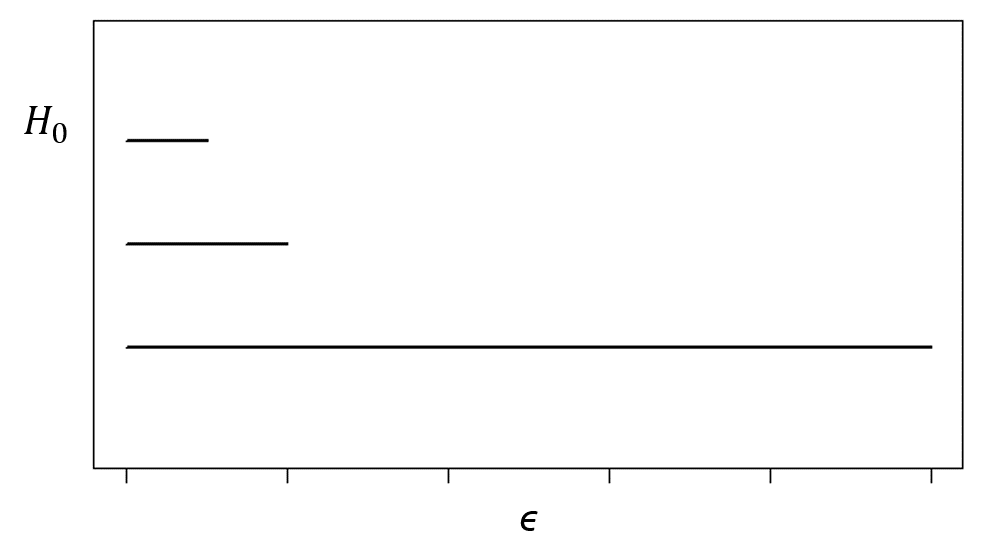}}
	\caption{Barcode for three qubit fully inseparable states. }
	\label{figure:3}
\end{figure}
\FloatBarrier
{ In summary all states in this class give barcodes of the same form of the one  depicted in Figure~\ref{figure:3}, which however may differ in the length of the first two lines depending on the values of $ E(i,j) $, $E(i,\{j,k\})$  and $E(j,\{i,k\})$ .}


\section{Classification of four qubit states using $\widetilde{D}(i,j)$}

\begin{figure}[htbp]
	\centerline{\includegraphics[scale=0.43]{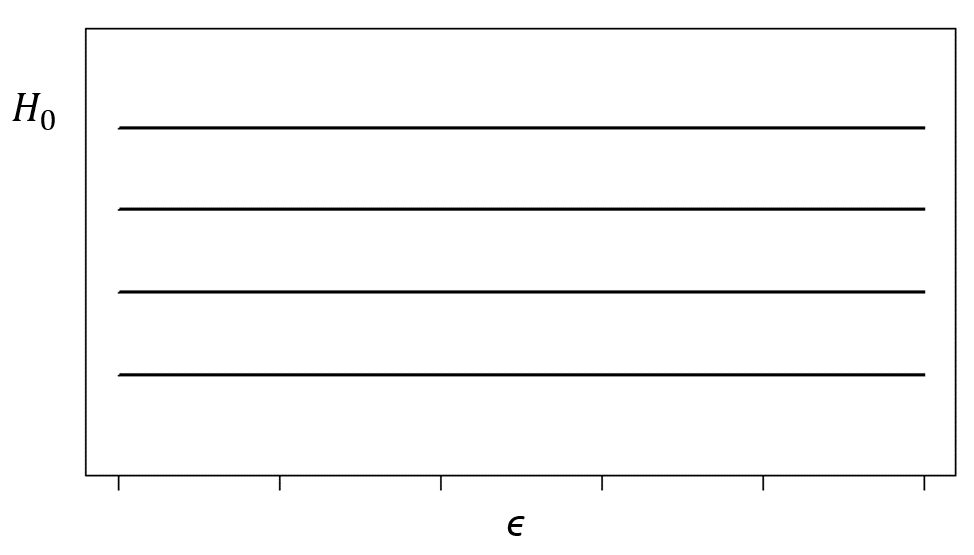}} 
	\caption{{ Barcode for product states of four qubits.}}
		\label{figure:4}
	\end{figure}

For four qubit states, Equation~\eqref{Dtilde} becomes
\begin{align}
\label{Dtilde4}
\widetilde{D}(i,j)=\Big[E(i,j)&+ E(i,\{j,k,l\}) E(j,\{i,k,l\}) \notag \\
&\times E(\{j,l\},\{i,k\})E(\{j,k\},\{i,l\})
\Big]^{-1}
\end{align}
with $i,\neq j\neq k\neq l$. 
In a similar way as shown for the three qubit case, it is possible to distinguish between the following classes, { each one associated to a different separability class}. 
\\\\
\noindent\emph{{ $\triangleright  $ Product states }}\\ { Pure states of this kind  are fully separable ($|\psi_1\rangle|\psi_2\rangle|\psi_3\rangle|\psi_4\rangle$) and they have associated} the barcode in 
	Figure~\ref{figure:4};
	\FloatBarrier
	
\begin{figure}[htbp]
	\centerline{\includegraphics[scale=0.43]{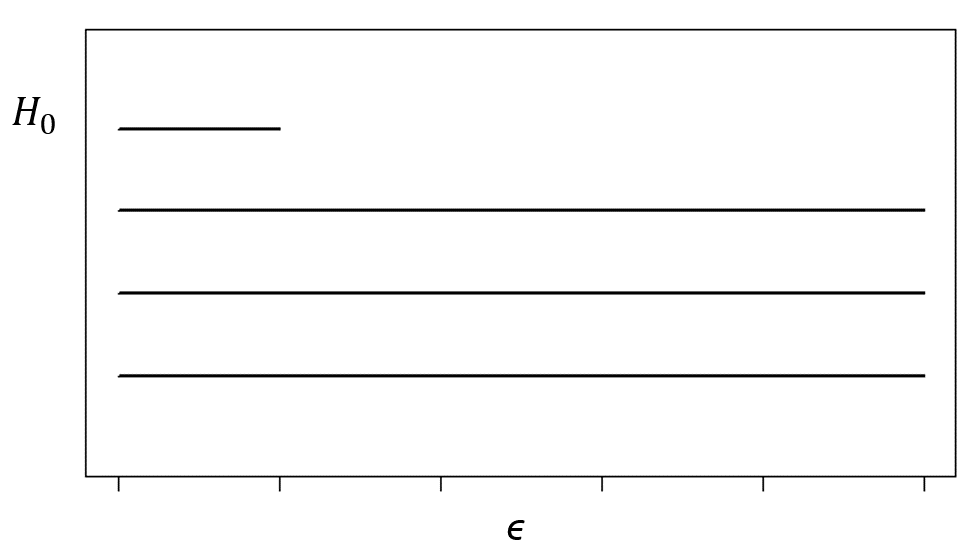}} \caption{Barcode for four qubit tri-separable  states. }
		\label{figure:5}
	\end{figure}

\noindent\emph{{ $\triangleright  $ Tri-separable  states }}\\
{ States that are tri-separable (i.e. of the form $|\psi_1\rangle|\psi_2\rangle|\psi_{34}\rangle$) }
		have associated the barcode in 
	Figure~\ref{figure:5};
	\FloatBarrier
		
\begin{figure}[htbp]
	\centerline{\includegraphics[scale=0.43]{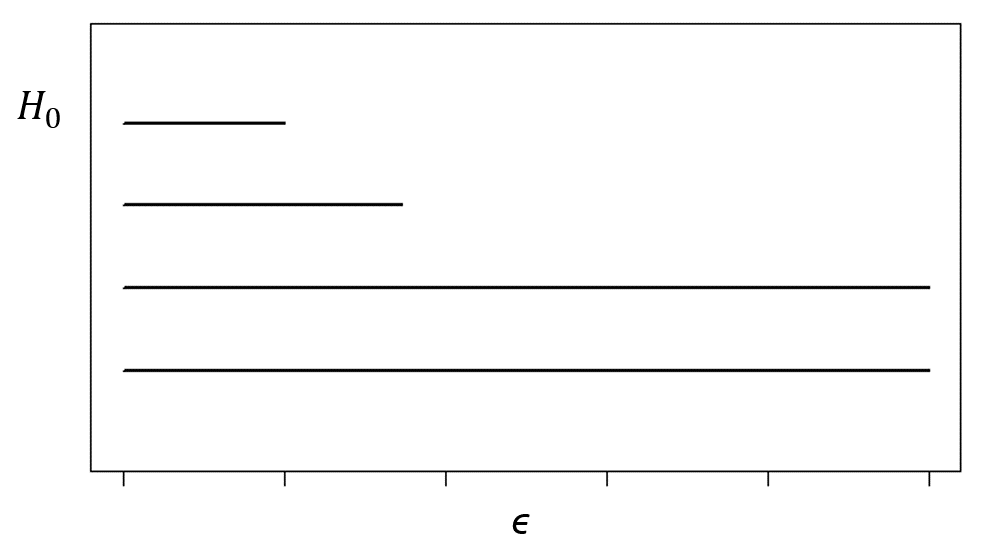}} 
	\caption{Barcode for four qubit bi-separable  states. 
}		
         \label{figure:6}
	\end{figure}

\begin{figure}[htbp]
		\centerline{\includegraphics[scale=0.42]{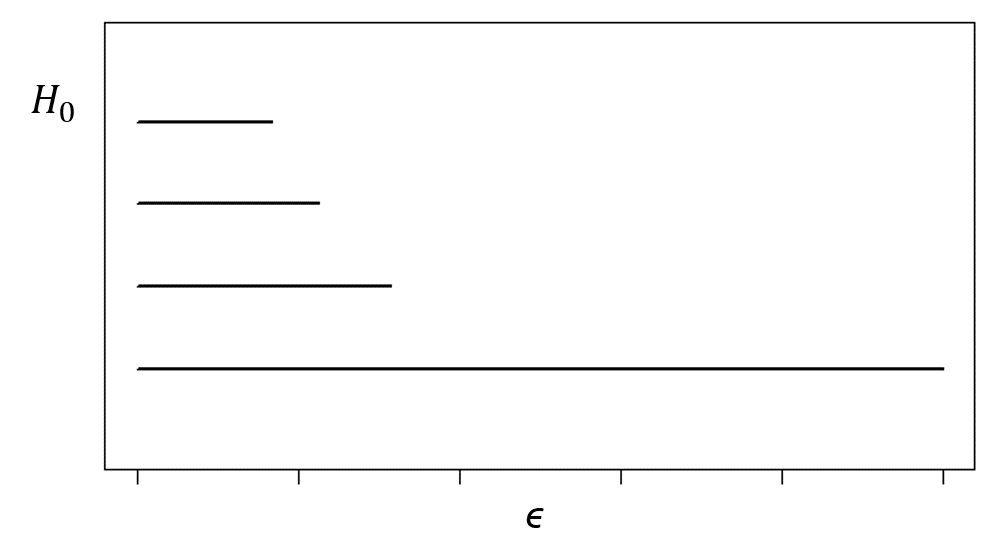}} 		\caption{Barcodes for true entangled states of  four qubits. 
	 }
		\label{figure:7}
	\end{figure}

	\noindent\emph{{ $\triangleright  $ Bi-separable  states }}\\ { States that are bi-separable (i.e. of the form  $|\psi_1\rangle|\psi_{234}\rangle$)} have associated the barcode in Figure~\ref{figure:6};
	\FloatBarrier
\noindent\emph{{ $\triangleright  $ Fully inseparable states}}\\
{ Such pure  states  are four-partite entangled  (i.e.  of the form $|\psi_{1234}\rangle$) and they are all mapped to  clouds of four points  at finite distance  from each another. 	States of this class  have associated  the  barcodes shown in Figure~\ref{figure:7}.}
\FloatBarrier

{ It is worth noticing that, for a generic  $n$-qubit quantum state, barcodes obtained  with distance (\ref{Dtilde})  are always able to distinguish between the standard  separability classes.  The  number of bi-partitions  needed to calculate $ \widetilde{D}(i,j) $ grows with the number of qubits as $2^{n-2}+1$.  However,  barcodes for the $0$-order homology group $ H_0 $ are sufficient to recover the standard separability classes, since the  number of  connected components appearing in the barcode is always upper bounded by the number of qubits $n$.  }


\section{{ Classification of fully inseparable qubit states}}

{ In the previous section we have provided  a classification  of multi-qubit states in  separability classes. We now concentrate on the fully inseparable set. We will refer to these states also as "true" or "genuine" entangled states.
It was shown in \cite{dur} that pure states of three-qubit systems showing tripartite entanglement can be further classified into two inequivalent classes under Stochastic Local Operations and Classical Communication (SLOCC).
By using the same SLOCC based approach, Verstraete et al.
\cite{Verstraete} classified generic 4-qubit quantum states  in nine different ways. 
In this section we will address  the problem  of whether or not a more refined classification of true entangled states  is possible by means of barcodes.   We will show that this can  be done by resorting to the distance $D$ defined by Equation~\eqref{D}.} \\\\

	\noindent\emph{{ $\triangleright  $ Three qubits case}}\\
{ In the case of genuine three-partite entangled states of three qubits  it is possible to recognize the  classes listed below. In order to visually distinguish them, we  draw the simplicial complex obtained from the point cloud { when  $\epsilon$ is larger  than the maximum distance  $ D(i,j) $ between  pairs of  qubit.} }


\begin{description}
\item[$a)$] This subclass collects all those states  where  entanglement between any possible pair of the three qubits gives $E(i,j)=0$, for all $ i,j$.
This implies that each point of our three points cloud is infinitely far away from the others. Hence they generate a barcode that only displays the 0-order homology group $H_0$, i.e the connected components as shown in Figure \ref{figure:8}.
\begin{figure}[htbp]
	\centerline{\includegraphics[scale=0.4]{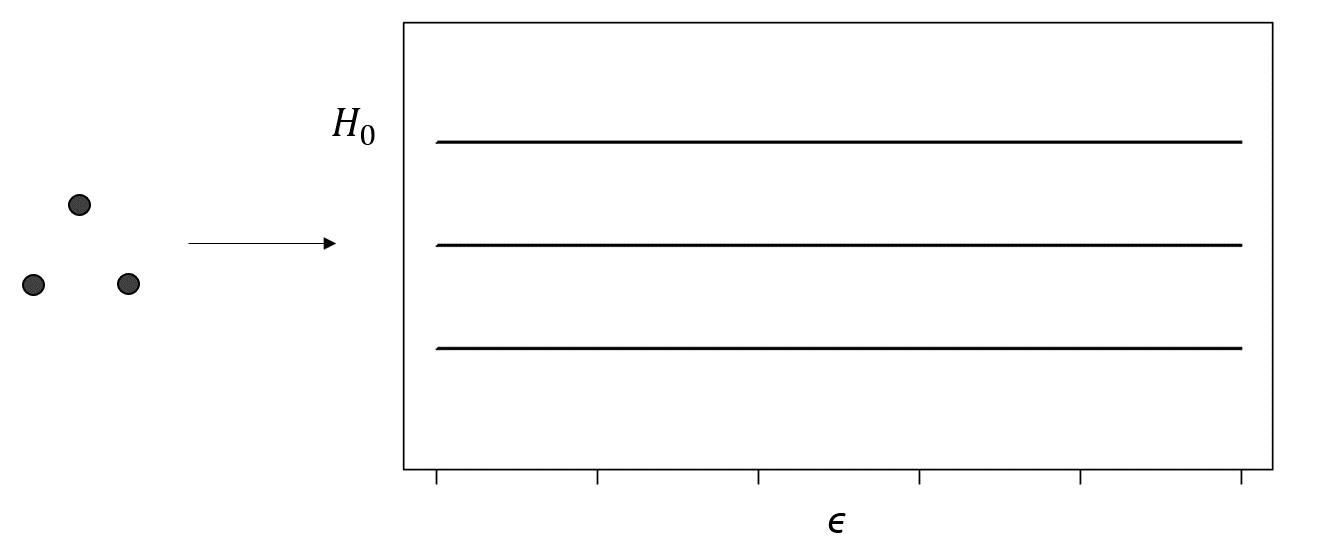}} 
	\caption{Barcode for three qubit states a) }
	\label{figure:8}
\end{figure}
\FloatBarrier
{ The state   $ \left| GHZ\right\rangle_3=\dfrac{1}{\sqrt{2}}\left( |000\rangle+|111\rangle\right)$ can be chosen to be a representative for this class, which exactly corresponds to the GHZ-class of  \cite{dur}. }

\item[$b)$] Another class is defined by states where $E(i,j)>0$,  $E(j,k)=0$ and $E(i,k)=0$; the associated point cloud is made of three points where the first and the second are at a finite distance, while the third point is at infinite distance from the other two.
The  barcode shows  three different  connected components in the interval $ \left[ 0,\frac{1}{E(i,j)}\right] $, while for values grater than $\frac{1}{E(i,j)}$ only two of them persist as shown in Figure~\ref{figure:9}.
\begin{figure}[htbp]
\centerline{\includegraphics[scale=0.4]{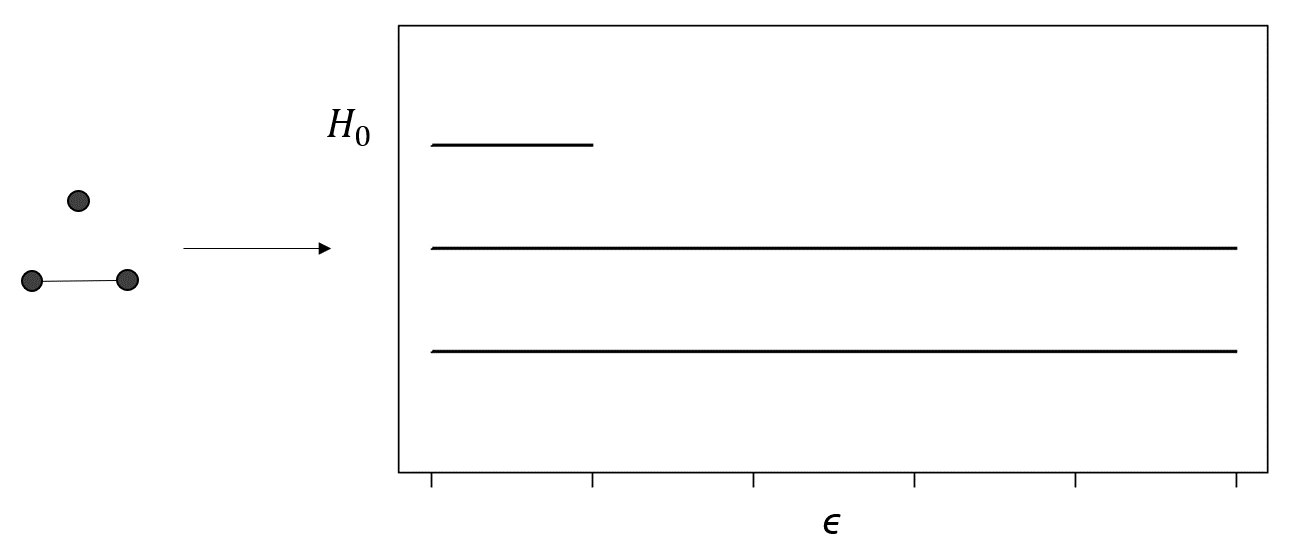}} 
\caption{Barcode for three qubit states b) }
	\label{figure:9}
\end{figure}
\FloatBarrier
{ The state   $ \left| \psi_b\right\rangle_3=\dfrac{1}{\sqrt{3}}\left( |000\rangle+|011\rangle+|111\rangle\right)$ can be chosen to be a representative for this class. Note that this class constitutes a subset of the W-class found under SLOCC \cite{dur}. }

\item[$c)$]{ This class contains the following states.\\
- States with  $E(i,j),E(j,k)>0$ and $E(i,k)=0$; their associated point cloud is made of three points where the first and the second are at distance $\frac{1}{E(i,j)} $, the second and the third are at distance $\frac{1}{E(j,k)} $, but the first and the third points are at infinite distance. This can be seen in the barcode of Figure~\ref{figure:10} where two of the three lines vanishes and only one persists. A representative for this class, with such properties could be the state  $ \left| \psi_c\right\rangle_3=\dfrac{1}{\sqrt{3}}\left( |000\rangle+|001\rangle+|011\rangle+|111\rangle\right)$. \\
-  States where, at a finite value $ \epsilon^* $ of $D(i,j)$, the three points get connected one with each other to form a triangular graph that is immediately filled with a 2-simplex. This means that for $ \epsilon\geq \epsilon^* $ we are left with only one connected component (the 2-simplex), again shown by the barcode of Figure~\ref{figure:10}. A representative for this class with such properties could be the state  $ \left| W\right\rangle_3=\dfrac{1}{\sqrt{3}}\left( |100\rangle+|010\rangle+|001\rangle\right)$.}
\begin{figure}[htbp]
	\centerline{\includegraphics[scale=0.4]{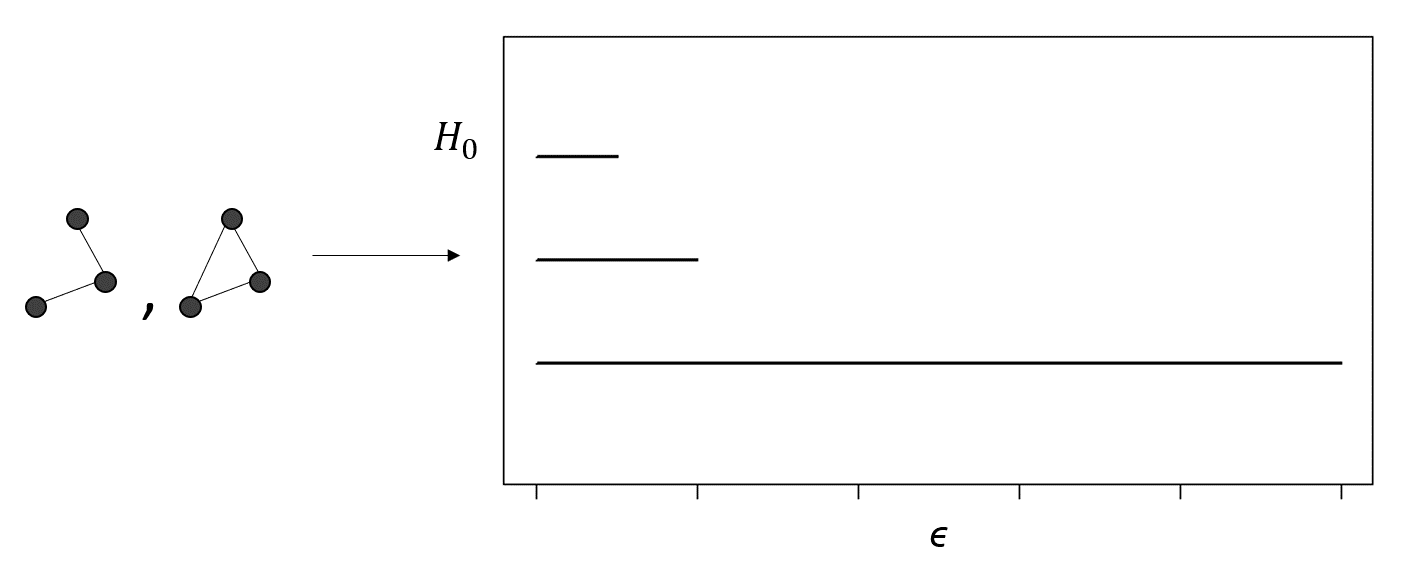}} \caption{Barcode for three qubit states c)}
	\label{figure:10}
\end{figure}
\FloatBarrier
{ States of this class constitute another subset of the W-class found using the SLOCC classification. More precisely, the classes b) and c) form a partition of the W-class. In the last section we will show that using another approach to  constructing the simplicial complex, it is possible to further distinguish $\left| W\right\rangle_3  $ from $ \left| \psi_c\right\rangle_3 $ states.}
\end{description}
{ In summary, we have shown that a classification performed using barcodes obtained with distance $ D(i,j) $ is able to distinguish   3 different classes of true entangled states of three qubits, hence going beyond the known SLOCC classification.}\\

\begin{figure}[htbp]
	\begin{center}
		\includegraphics[scale=0.37]{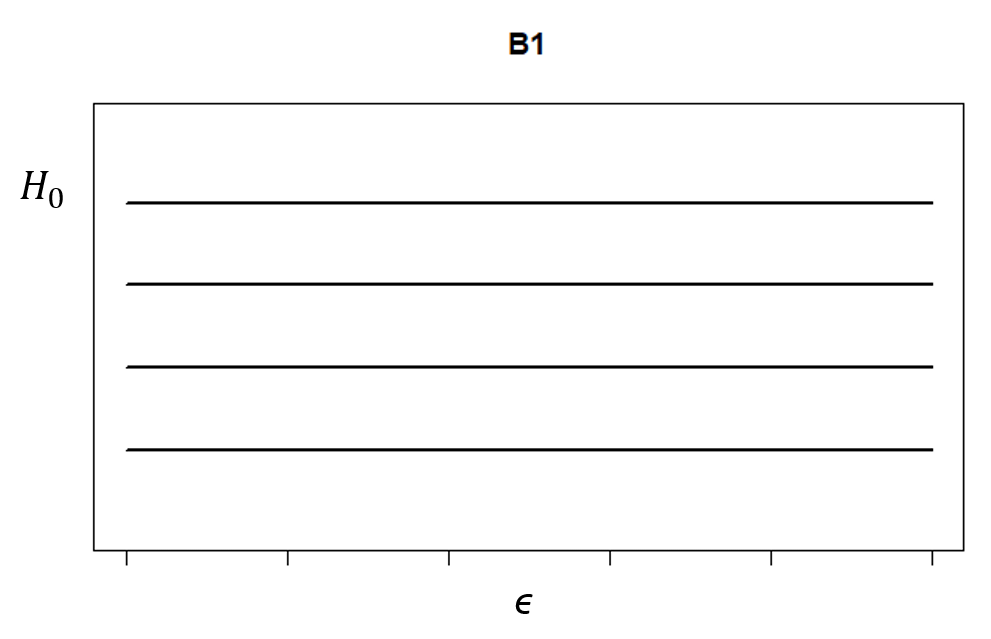}\\
		\includegraphics[scale=0.37]{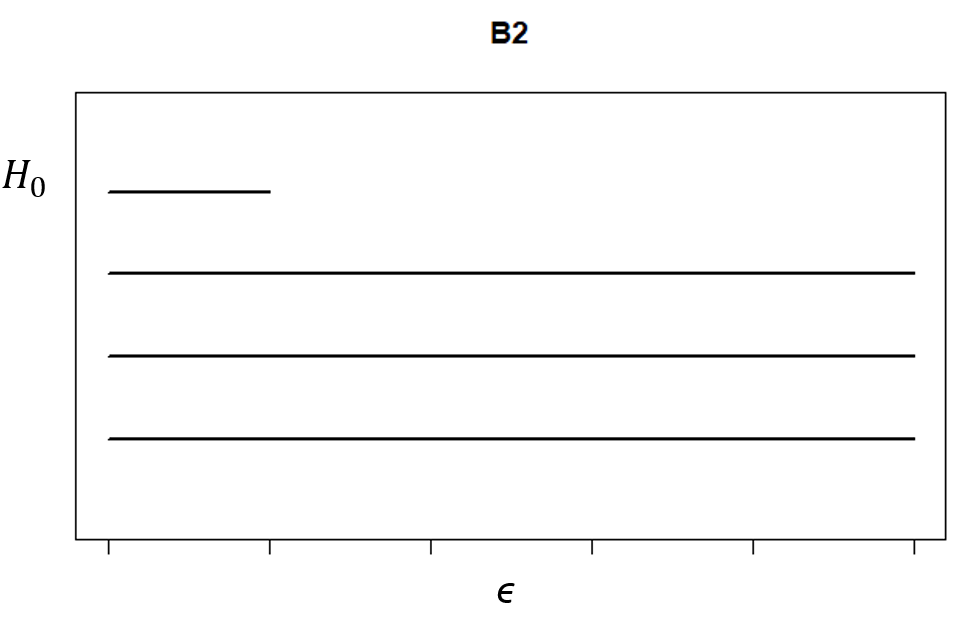}\\
		\includegraphics[scale=0.37]{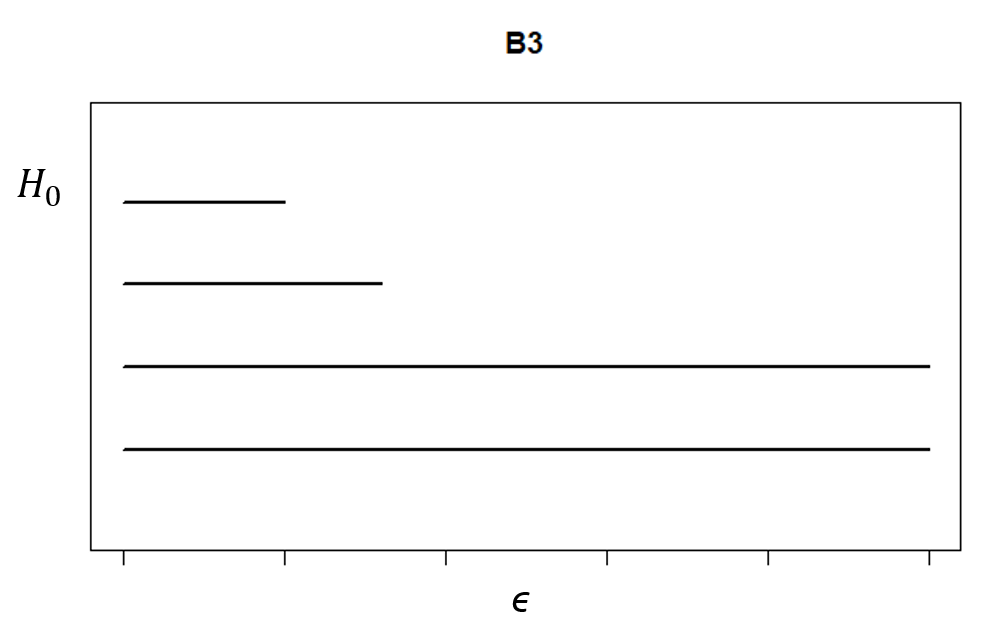}\\
		\includegraphics[scale=0.37]{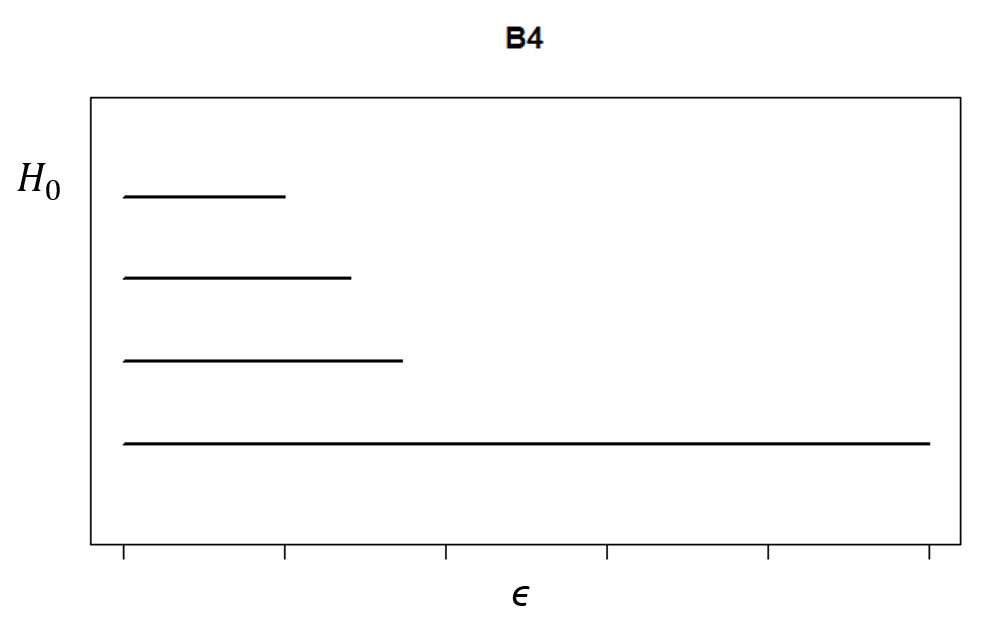}
		\includegraphics[scale=0.37]{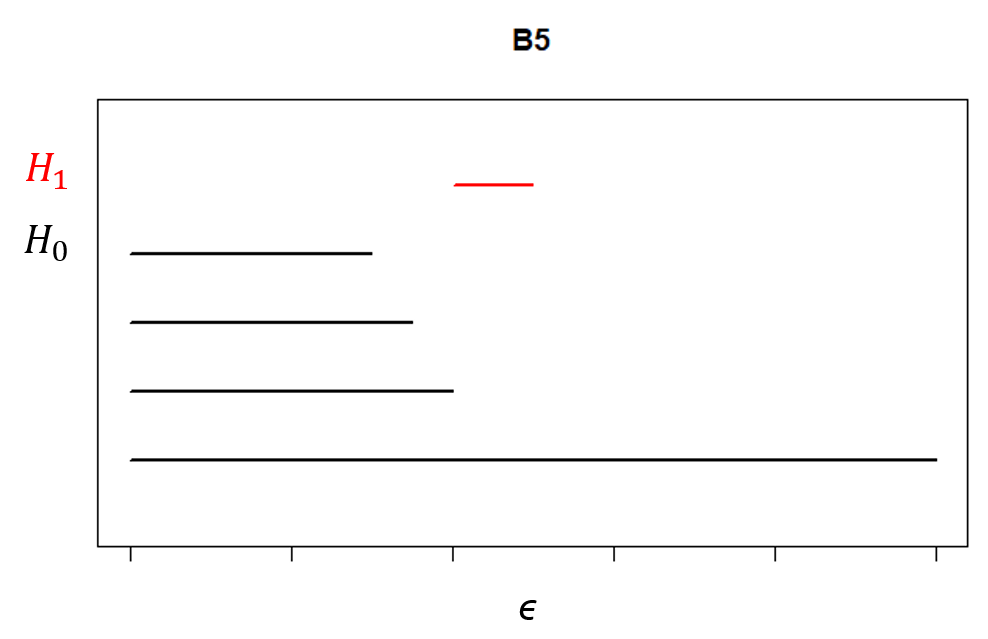}\\
		\includegraphics[scale=0.37]{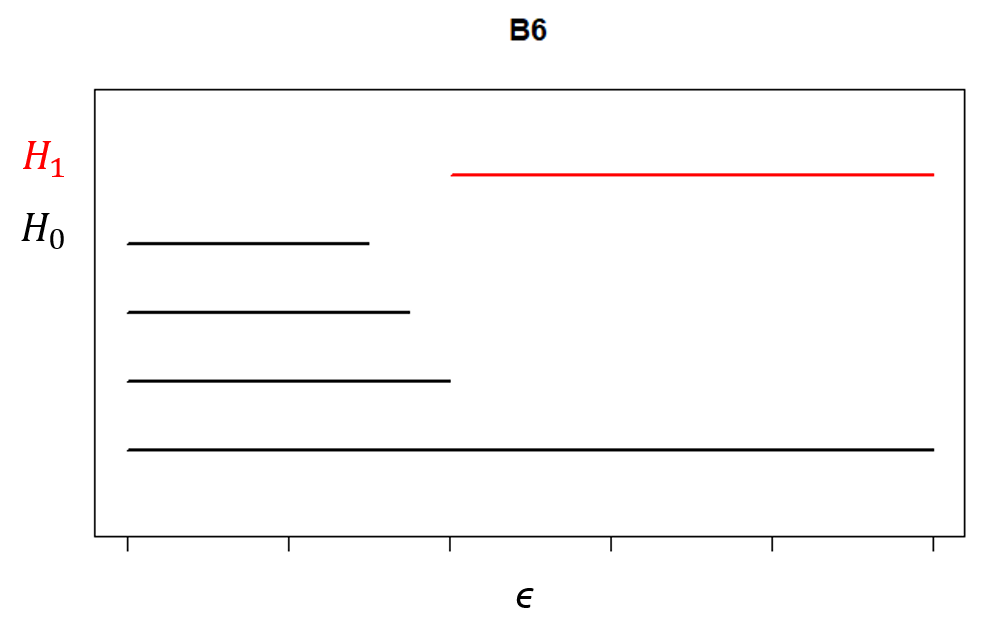}  
		\caption{Four qubit barcodes B1-B6}
		\label{figure:11}
	\end{center}
\end{figure}

\noindent\emph{{ $\triangleright  $ Four qubits case}}\\
{ In the case of four qubit states with genuine four-partite entanglement we can identify six different classes.} The results obtained using the distance \eqref{D} are summarised in Table~\ref{tab1}, which refers to barcodes of Figure~\ref{figure:11}. { Also in this case we use  the simplicial complex obtained when  $\epsilon$ is larger  than the maximum distance  $ D(i,j) $ between  pairs of  qubit in order to visually distinguish different states. }  

\FloatBarrier
\begin{table}[htbp]
	\begin{tabular}{ |c| c | c | p{4cm} | }
		\hline
Class & \small Barcode & \small Complex  &  \small Representative State    \\ \hline
&  & &  \\ \hline
a)& B1   & \raisebox{-1\height}{\includegraphics[scale=0.5]{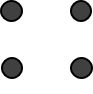}} &  $|GHZ\rangle_{4} $   \\ \hline
	b) & B2  &  \raisebox{-1\height}{\includegraphics[scale=0.5]{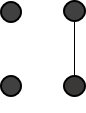}} &   \footnotesize$|B\rangle=|0000\rangle+|0111\rangle+|1101\rangle $  \\ \hline
	c)&	B3 & \raisebox{-1\height}{\includegraphics[scale=0.5]{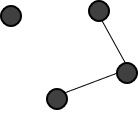}} &   \footnotesize$|C\rangle=|0000\rangle+|0111\rangle+|1010\rangle++|1011\rangle $   \\ \hline
	$ $&	B3 &  \raisebox{-1\height}{\includegraphics[scale=0.5]{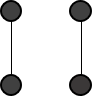}} &  \footnotesize$|C'\rangle=|0000\rangle+|0011\rangle+|1111\rangle $    \\ \hline
		$ $&	B3 & \raisebox{-1\height}{\includegraphics[scale=0.5]{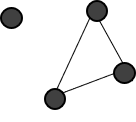}} & \footnotesize$|C''\rangle=|0011\rangle+|1011\rangle++|1101\rangle+|1110\rangle $    \\ \hline
		d) &	B4 & \raisebox{-1\height}{\includegraphics[scale=0.5]{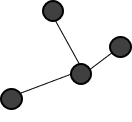}} & \footnotesize $|D\rangle=|0000\rangle+|0001\rangle+|1001\rangle+$  $+|1101\rangle+|1011\rangle+|1111\rangle $    \\ \hline
		$ $&	B4 & \raisebox{-1\height}{\includegraphics[scale=0.5]{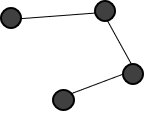}} &   \footnotesize$|D'\rangle=|0000\rangle+|0011\rangle++|0111\rangle+|1110\rangle+|1111\rangle $   \\ \hline
		$ $&	B4 & \raisebox{-1\height}{\includegraphics[scale=0.5]{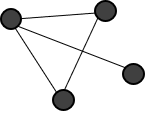}} &  \footnotesize $|D''\rangle=|0000\rangle+|0011\rangle++|0110\rangle+|0111\rangle+|1011\rangle $    \\ \hline
		$ $&	B4 & \raisebox{-1\height}{\includegraphics[scale=0.5]{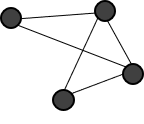}} & \footnotesize $|D'''\rangle=|0000\rangle+|0001\rangle++|0011\rangle+|0101\rangle+|0111\rangle++|1101\rangle  $   \\ \hline
		$ $&	B4  & \raisebox{-1\height}{\includegraphics[scale=0.5]{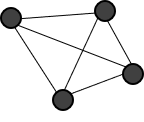}} & \footnotesize$|W\rangle_{4} $\\ \hline
		e) &	B5 & \raisebox{-1\height}{\includegraphics[scale=0.5]{4k}} &\footnotesize $|E\rangle=2|0000\rangle+|0011\rangle++|0110\rangle+|1001\rangle+|1100\rangle++2|1111\rangle   $    \\ \hline
		f) &	B6 & \raisebox{-1\height}{\includegraphics[scale=0.5]{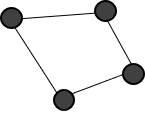}} &  \footnotesize$|F\rangle=|0000\rangle+|0011\rangle+|1010\rangle++|1111\rangle  $   \\ \hline
	\end{tabular}		
\caption{Classification of  four-partite entangled states of four qubit.}	
\label{tab1}
\end{table}
\FloatBarrier
{ As we can see from Table~\ref*{tab1}, both classes c) and d)  contain states that share the same homology groups although but have different properties (cf. barcodes B3 and B4 in Figure \ref{figure:11}).
	Moreover, it is worth noticing that states like $|W\rangle_{4} $  and those in class e) are both represented  by a 3-simplex for  sufficiently large $ \epsilon$.    The  difference  is due to the fact that the barcode of states in e) show a hole, i.e. homology group $ H_1 $ depicted in red in barcode B5, for a short interval. For large   $ \epsilon $, this hole gets filled with a  3-simplex and disappears.

 As previously said for the three qubit case, we will show in the following section how it is possible  to refine also  the  classification of four qubit entangled states by using a different kind of complex. }

\section{Rips vs \v{C}ech complex}
\label{Cech}

\begin{figure}[htbp]
	\centerline{\includegraphics[scale=0.43
		]{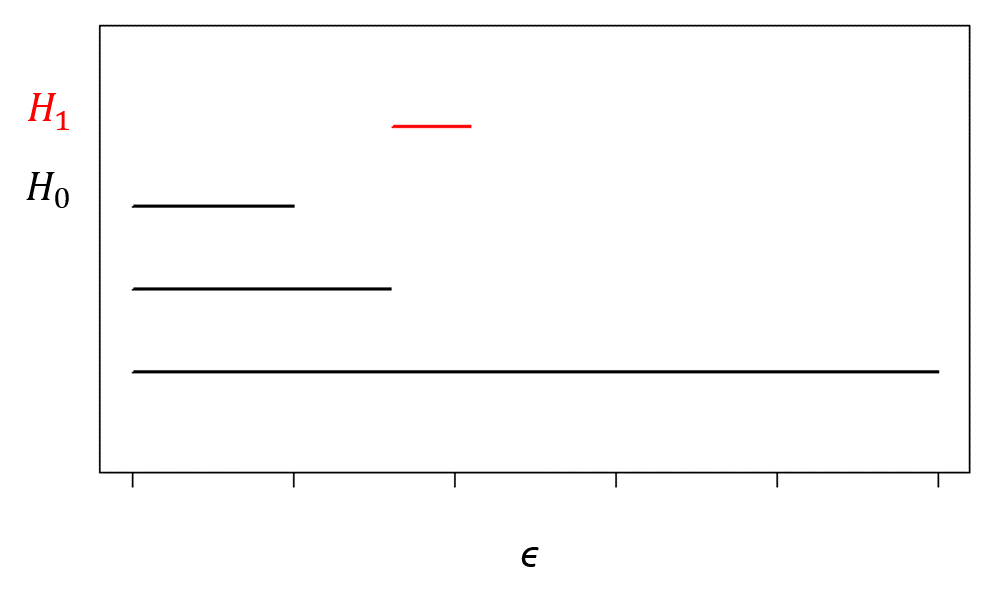}} 
	\caption{Barcode for W-like states 
		with \v{C}ech  complex. }
	\label{figure:12}
\end{figure}

\begin{figure}[htbp]
	\vspace{-0.84cm}
	\begin{center}
			\includegraphics[scale=0.39]{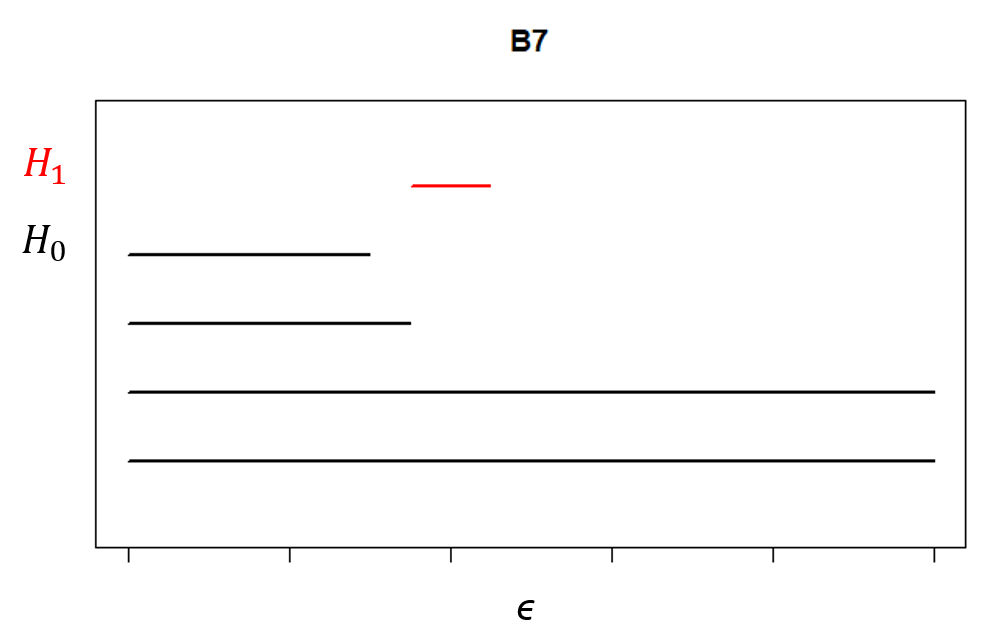}
	\includegraphics[scale=0.39]{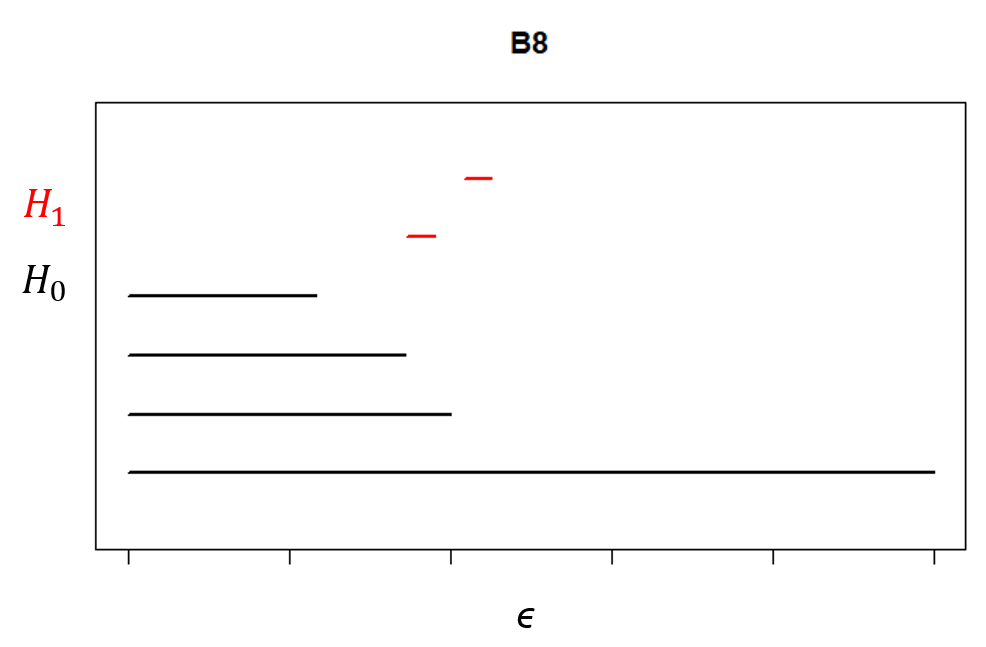}
	\includegraphics[scale=0.39]{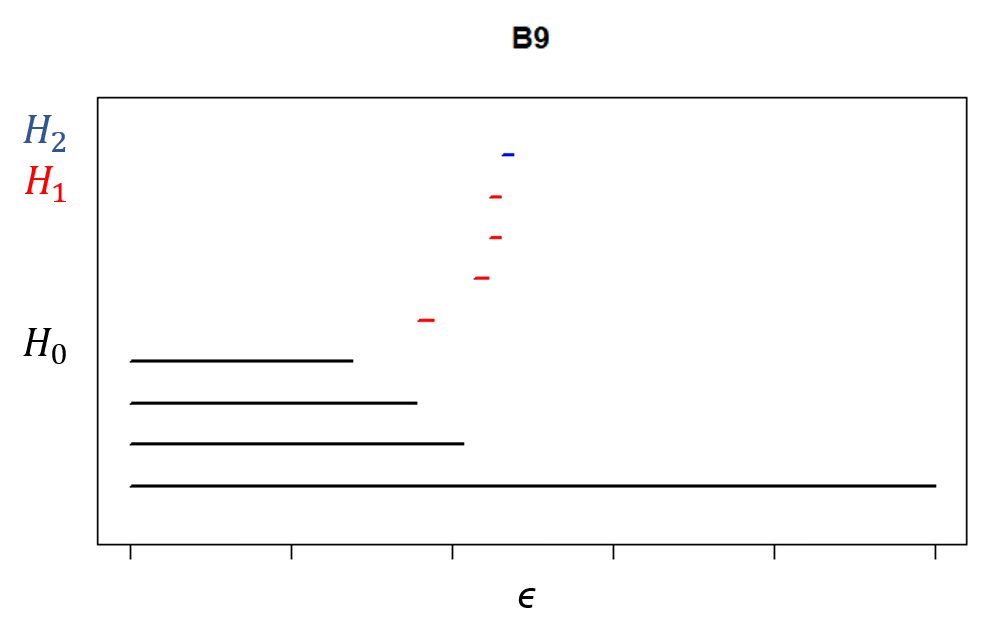} 
	\includegraphics[scale=0.34]{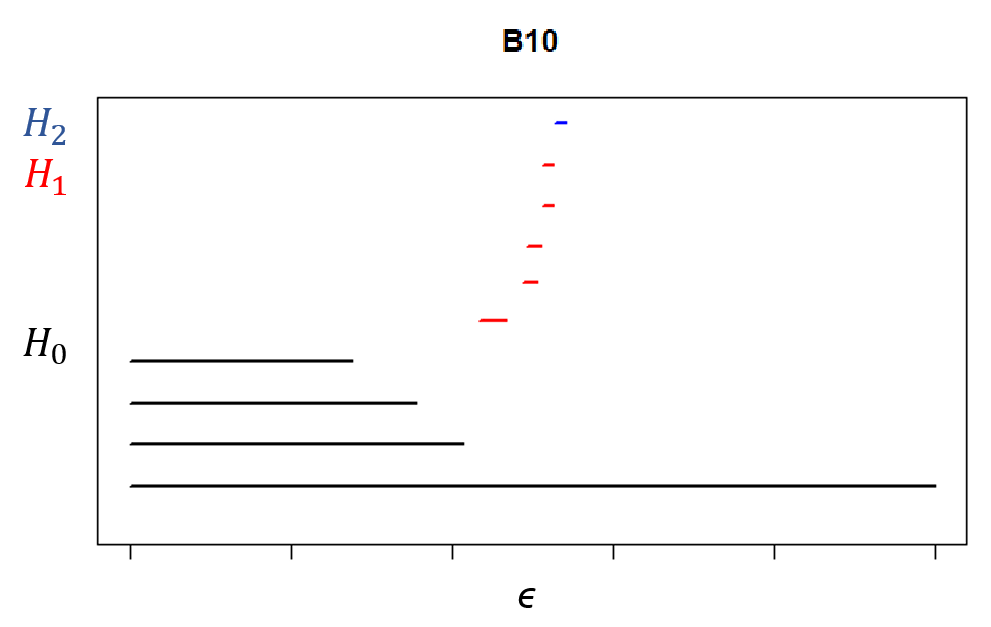} 
	\caption{Four qubit barcode B7-B10. }
	\label{figure:13}
		\end{center}
\end{figure}

{The Rips complex is closey related to another simplicial complex, called the \v{C}ech complex. This is defined on a set of balls and  has a simplex for every finite subset of balls with nonempty intersection; thus, while in Rips complexes, $k$-simplices  correspond to $(k + 1)$ points which are pairwise within distance $ \epsilon$, in \v{C}ech complexes $k$-simplices are determined by $(k + 1)$ points  whose closed 
$ \epsilon/2$-ball neighborhood have a point of common intersection \cite{Barcodes}.

By the \v{C}ech theorem \cite{Borsuk}, the \v{C}ech complex has the same topological structure as the open sets ($ \epsilon$-balls) cover of the point cloud. This is not true for the Rips complex, which is  more coarse than the \v{C}ech complex. Therefore, the latter is a more powerful tool for classification with respect to Rips. In fact, it is possible to refine the  classification of the fully inseparable three qubit states in class c). 
The reason why this is possible is that in the \v{C}ech complex, triangle graphs like the one that appears in the W-state point cloud (Figure~\ref{figure:10}) are not immediately filled with a 2-simplex. In fact a hole  (1-order homology group $H_1$) appears for a short interval of the values of $ \epsilon $  before the graph gets filled with the 2-simplex.

{ Therefore, for the c) class in the three qubit case,  barcodes associated to states like $ \left| \psi_c\right\rangle_3 $ remain the same as in Figure~\ref{figure:10}, the one for the W-like states instead becomes as shown  in Figure~\ref{figure:12}.}}
\FloatBarrier
{ Analogously, in the four qubit case, the \v{C}ech complex allows us to refine the classification of
states in c) and d) classes of Table~\ref{tab1}. This is shown   in  Table~\ref{tab2} with reference to the new barcodes of Figure~\ref{figure:13}:}
\begin{table}[htbp]
\begin{center}
	\begin{tabular}{ |c |c |c |}
		\hline
		\v{C}ech barcode & Rips barcode  & State  \\ \hline
	 B1 &  B1 & $ |GHZ\rangle_{4} $ \\ \hline
		 B2 &  B2 & $ |B\rangle $ \\ \hline
	B3 &  B3 & $ |C\rangle $  \\ \hline
	B3 &  B3 & $ |C'\rangle $  \\ \hline
			B4	 &  B4 & $ |D\rangle $  \\ \hline
		B4 &   B4 &$ |D'\rangle $  \\ \hline
	B5	 &   B4 & $ |D''\rangle $  \\ \hline
B6	 &   B6 & $ |F\rangle $ \\ \hline
	B7	 &  B3 & $ |C''\rangle $  \\ \hline
	B8	 &   B4 & $ |D'''\rangle $ \\ \hline
	B9 &  B4 & $ |W\rangle_4 $  \\ \hline
		B10 &  B5 & $ |E\rangle $   \\ \hline
	\end{tabular}	
	\caption{Comparison between barcodes obtained from the Rips and \v{C}ech complex for fully inseparable states of four qubits}
	\label{tab2}		
\end{center}
\end{table}
\FloatBarrier

\section{Comparison with SLOCC and generalisation}
\label{comparison}

{ In \cite{Verstraete} it is proposed a classification for pure entangled states of four qubit. This classification is   obtained from  the orbits generated by SLOCC operations and produces nine classes. Among these nine classes, one is called 'generic' and contains an uncountable number of SLOCC inequivalent classes of states, as shown in \cite{Gour}.
A comparison between the classification based on persistent homology and the one presented in \cite{dur} is possible in the three qubit case, as discussed before. However, in the four qubit case, a comparison between our approach and the one in \cite{Verstraete} does not allow us to clearly establish a correspondence between classes as in the three qubit case.
In order to better understand this, it is useful to consider few specific examples.
Consider the class \\
$ L_{ab_3}=a(|0000\rangle+|1111\rangle)+\frac{a+b}{2}(|0101\rangle+|1010\rangle)++\frac{a-b}{2}(|0110\rangle+|1001\rangle)+\frac{i}{\sqrt{2}}(|0001\rangle+|0010\rangle+|0111\rangle++|1011\rangle)  $ 
 defined in \cite{Verstraete}. It is easy to check that for the different values of the parameters $ a $ and $ b $ we obtain the following states:\\
- for $ a=b=0 $, a W-like state with associated barcode B4;\\
-for $ a=0 $ and $  b=1 $, a state with  barcode B6;\\
-for $ a=b=1 $, a state with   barcode B3.\\
Moreover states belonging to the class $ L_{0_{7\oplus\bar{1}}}=|0000\rangle+|1011\rangle+|1101\rangle+|1110\rangle $ have the same barcode of  $ |GHZ\rangle_4 $ which instead belongs to the generic class $ G_{abcd} $  for   $ a=d $ and $ b=c=0 $.  
This makes it impossible to fully characterize  the SLOCC classification proposed in \cite{Verstraete} in terms of  barcodes  and vice versa.
The SLOCC approach is indeed based on a classification criteria (equivalence with respect to local operations) that are not comparable ours, which are of different nature (the change in the topology of the point cloud formed by the qubit, depending on the pairwise entanglement strength).
Therefore the classifications obtained are intrinsically different besides leading  to  a different number of classes: nine with SLOCC and six (using Rips) or ten (using \v{C}ech) with persistent homology.
We will nevertheless argue in the following that our approach is mor robust in terms of increasing the number of qubits. In fact, it  was shown in \cite{dur} that  for systems of size $ n\geq 4 $ there exist infinitely many inequivalent kinds of entanglement  under SLOCC and no finite classification is known  for states of those sizes.  If we restrict to the case of genuine multipartite entangled state of $n$ qubit, it is easy to see that our persistent homology approach always provides a finite number of different classes for any value of $n$.
This is due to the fact that the total number of possible homology groups  that can be obtained considering a data set of n point is always finite. The number of possible  barcodes $ B_n $ obtained  with a data set of $n$ points can in fact be bounded from the above as follows
\begin{equation}
B_n<\left( \sum_{e=0}^{\frac{n(n-1)}{2}}G_n(e)\ e!\right) 
\end{equation}
where   $ G_n(e) $ is the total number of possible graphs with $ n $ vertices and $ e $ edges, up to permutations of the vertices, and the factorial $ e! $ takes into account all the possible ways of constructing  $ G_n(e) $.

\section{Conclusion}

We have proposed the use of  persistent homologies for the study of multi-partite entanglement.
The analysis we have shown, although being a topological analysis gathering as such  qualitative information, turns out to be quite powerful. 
A comparison between the persistent homology classification and   known  SLOCC classifications \cite{dur,Verstraete}  highlighted the  different nature of these classifications.
While on one hand, due to the non local behaviour of entanglement,   it is reasonable to use local quantum operations to find equivalence classes, on the other hand, topological properties of entanglement   that can be visualised with  barcodes are studied. 
Choosing  $ \widetilde{D} $ as a distance allows us to     represent separability  properties of a given state.
Distance $ D $ instead has been applied to  genuinely entangled states to give insights about the topology arising from the bipartite entanglement between pairs of qubits.
This analysis allowed us to idenitfy states displaying the same homology groups, providing a new classification.

Looking ahead we plan to apply the presented approach 
to genuine multipartite entanglement in the case of 5 qubits.
There, having no a priori information on entangled states structures
we will randomly generate entangled states, then analyse 
their barcodes, and on the basis of these latter 
We will single out possible entanglement classes.  
This method would be efficient when also extended to 
very large quantum data sets since Eq.(\ref{D}) implies the calculation 
of $ n(n-1)/2 $ correlation measures [in contrast, the classification of 
all states including partially and fully separable would be of 
exponential complexity according to Eq.(\ref{Dtilde})]. 
Future investigations should also seek 
true metrics (rather than just semi-metrics) on the quantum
data set,  since triangle inequality ensures   the point cloud  to belong to an Euclidean space.}






\begin{thebibliography}{99}

\bibitem{Horo}
{Horodecki R., et al.}
{Rev. Mod. Phys.}{ 81}{2009}{865}.

\bibitem{Carl}
{Carlsson G. and  Zomorodian A.}
{Discrete Comput. Geom.}{ 33}{ 2005}{249}.

\bibitem{Edel}
{Edelsbrunner H., Letscher D., and Zomorodian A.}
{Discrete Comput. Geom.}{28}{ 2002}{511}.

\bibitem{Seth}
{Lloyd S., Garnerone S., and Zanardi P.}
{arXiv:1408.3106 [quant-ph]}{}{ 2014}{}.

\bibitem{Hatcher}
{Hatcher, A.}
{Cambridge University Press.}{}{2002}{}.

\bibitem{Carl2}
{Carlsson, E. et al.}
{Int. J. Comput. Geom. Appl.}{16}{2006}{291}.

\bibitem{Borsuk}
{Borsuk K.}
{Fundamenta Mathematicae}{35(1):217--234}{1948}.


\bibitem{Barcodes}
{ Ghrist, R.}
{Bulletin of the American Mathematical Society, Vol. 45}{}{2008}{}
{
\bibitem{dur}
{  Dur W., Vidal G., and Cirac J. I.}
{Phys. Rev. A 62, 062314}{}{2000}{}

\bibitem{Verstraete}
{  Verstraete F.,  Dehaene J.,  De Moor B., and  Verschelde H.}
{Phys. Rev. A 65, 052112}{}{2002}{}

\bibitem{Gour}
{ G. Gour, N. R. Wallach}
{Journal of Mathematical Physics 51, 112201 }{}{2010}{}

}


\end{thebibliography}
\end{document}